# LOFT - the Large Observatory For x-ray Timing

M. Feroci[1,1b,*]; J.W. den Herder[2]; E. Bozzo[3]; D. Barret[8]; S. Brandt[19]; M. Hernanz[6]; M. van der Klis[5]; M. Pohl[25]; A. Santangelo[17]; L. Stella[27]; A. Watts[5]; J. Wilms[87]; S. Zane[29]; M. Ahangarianabhari[14]; A. Alpar[4]; D. Altamirano[5]; L. Alvarez[6]; L. Amati[7]; C. Amoros[8]; N. Andersson[9]; A. Antonelli[10]; A. Argan[1]; R. Artigue[8]; P. Azzarello[3]; G. Baldazzi[103], S. Balman[11]; M. Barbera[12,106]; T. Belloni[13]; G. Bertuccio[14]; S. Bianchi[59]; A. Bianchini[15]; P. Bodin[114]; J.-M. Bonnet Bidaud[16]; S. Boutloukos[17]; J. Braga[18]; E. Brown[49]; N. Bucciantini[21]; L. Burderi[22]; M. Bursa[23]; C. Budtz-Jørgensen[19]; E. Cackett[111]; F.R. Cadoux[25]; P. Cais[26]; G.A. Caliandro[6]; R. Campana[1,1b]; S. Campana[13]; P. Casella[27]; D. Chakrabarty[28]; J. Chenevez[19]; J. Coker[29]; R. Cole[29]; A. Collura[106], T. Courvoisier[3]; A. Cros[8]; A. Cumming[31]; G. Cusumano[57], A. D'Aì[12]; V. D'Elia[10]; E. Del Monte[1,1b]; D. De Martino[31]; A. De Rosa[1]; S. Di Cosimo[1]; S. Diebold[17]; T. Di Salvo[12]; I. Donnarumma[1]; A. Drago[32]; M. Durant[33]; D. Emmanoulopoulos[107]; Y. Evangelista[1,1b]; A. Fabian[24]; M. Falanga[34]; Y. Favre[25]; C. Feldman[35]; C. Ferrigno[3]; M. H. Finger[36]; G.W. Fraser[35]; F. Fuschino[7]; D.K. Galloway[37]; J.L. Galvez Sanchez[6]; E. Garcia-Berro[6]; B. Gendre[10]; S. Gezari[62]; A.B. Giles[39]; M. Gilfanov[40]; P. Giommi[10]; G. Giovannini[102]; M. Giroletti[102]; A. Goldwurm[105]; D. Götz[16]; C. Gouiffes[16]; M. Grassi[56]; P. Groot[42]; C. Guidorzi[32]; D. Haas[2]; F. Hansen[19]; D.H. Hartmann[42]; C.A. Haswell[43]; A. Heger[37]; J. Homan[28]; A. Hornstrup[19]; R. Hudec[23,72]; J. Huovelin[45]; A. Ingram[46]; J.J.M. in't Zand[2]; J.Isern[6]; G. Israel[27]; L. Izzo[47]; P. Jonker[2]; P. Kaaret[49]; V. Karas[23]; D. Karelin[6]; D. Kataria[29]; L. Keek[49]; T. Kennedy[29]; D. Klochkov[17]; W. Kluzniak[50]; K. Kokkotas[17]; S. Korpela[45]; C. Kouveliotou[51]; I. Kreykenbohm[87]; L.M. Kuiper[2]; I. Kuvvetli[19]; C. Labanti[7]; D. Lai[52]; F.K. Lamb[53]; F. Lebrun[105]; D. Lin[8]; D. Linder[29]; G. Lodato[54]; F. Longo[55]; N. Lund[19]; T.J. Maccarone[107]; D. Macera[14], D. Maier[17]; P. Malcovati[56]; V. Mangano[57]; A. Manousakis[3]; M. Marisaldi[7]; A. Markowitz[109]; A. Martindale[35]; G. Matt[59]; I.M. McHardy[107]; A. Melatos[60]; M. Mendez[61]; S. Migliari[20]; R. Mignani[29,108]; M.C. Miller[62]; J.M. Miller[20]; T. Mineo[57]; G. Miniutti[112]; S. Morsink[64]; C. Motch[65]; S. Motta[13]; M. Mouchet[66]; F. Muleri[1,1b]; A.J. Norton[43]; M. Nowak[28]; P. O'Brien[35]; M. Orienti[102]; M. Orio[99,110]; M. Orlandini[7]; P. Orleanski[68]; J.P. Osborne[35]; R. Osten[69]; F. Ozel[70]; L. Pacciani[1,1b]; A. Papitto[6]; B. Paul[71]; E. Perinati[17]; V. Petracek[72]; J. Portell[6]; J. Poutanen[73]; D. Psaltis[70]; D. Rambaud[8]; G. Ramsay[76]; M. Rapisarda[1,1b]; A. Rachevski[77]; P.S. Ray[78]; N. Rea[6]; S. Reddy[80]; P. Reig[113,81]; M. Reina Aranda[63]; R. Remillard[28]; C. Reynolds[62]; P. Rodríguez-Gil[82,104]; J. Rodriguez[16]; P. Romano[57]; E.M.R. Rossi[83]; F. Ryde[84]; L. Sabau-Graziati[63]; G. Sala[6]; R. Salvaterra[85]; A. Sanna[61]; S. Schanne[16]; J. Schee[86]; C. Schmid[87]; A. Schwenk[88]; A.D. Schwope[89]; J.-Y. Seyler[114]; A. Shearer[90]; A. Smith[29]; D.M. Smith[58]; P.J. Smith[29]; V. Sochora[23]; P. Soffitta[1]; P. Soleri[61]; B. Stappers[91]; B. Stelzer[57]; N. Stergioulas[92]; G. Stratta[10]; T.E. Strohmayer[93]; Z. Stuchlik[86]; S. Suchy[17]; V. Sulemainov[17]; T. Takahashi[94]; F. Tamburini[15]; C. Tenzer[17]; L. Tolos[6]; G. Torok[86]; J.M. Torrejon[95]; D.F. Torres[96]; A. Tramacere[3]; A. Trois[1]; S. Turriziani[101], P. Uter[17]; P. Uttley[5]; A. Vacchi[77]; P. Varniere[105]; S. Vaughan[35]; S. Vercellone[57]; V. Vrba[97]; D. Walton[29]; S. Watanabe[94]; R. Wawrzaszek[68]; N. Webb[8]; N. Weinberg[28]; H. Wende[17]; P. Wheatley[98]; R. Wijers[5]; R. Wijnands[5]; M. Wille[87]; C.A. Wilson-Hodge[44]; B. Winter[29]; K. Wood[78]; G. Zampa[77]; N. Zampa[77]; L. Zampieri[99]; A. Zdziarski[50]; B. Zhang[100].

[1]IAPS-INAF, Via del Fosso del Cavaliere 100 - 00133 Rome, Italy; [1b]INFN, Sez. Roma Tor Vergata, Via della Ricerca Scientifica 1 - 00133 Rome, Italy; [2]SRON, Sorbonnelaan 2 - 3584 CA Utrecht, The Netherlands; [3]ISDC, Geneve University, Chemin d'Ecogia 16 - 1290 Versoix, Switzerland; [4]Sabanci University, Orhanli-Tuzla 34956, Istanbul, Turkey; [5]Astronomical Institute Anton Pannekoek, University of Amsterdam, Science Park 904 - 1098 XH Amsterdam, The Netherlands; [6]IEEC-CSIC-UPC-UB, Carrer del Gran Capità, 2 - 08034 Barcelona, Spain; [7]INAF-IASF-Bologna, Via P. Gobetti, 101 - 40129 Bologna, Italy; [8]IRAP, avenue du Colonel Roche, 9 - BP 44346 Toulouse, France; [9]Faculty of Physical and Applied Sciences, University of Southampton, Southampton, SO17 1BJ, United Kingdom; [10]ASDC, via Galileo Galilei - 00044 Frascati, Italy; [11]Middle East Technical University, Ankara, Mah. Dumlupınar Blv. No:1 - 06800 Çankaya Ankara, Turkey; [12]Dipartimento di Fisica, Palermo University, Via Archirafi, 36 - 90123 Palermo, Italy; [13]INAF-OA Brera, Via E. Bianchi 46 – 23807 Merate (LC), Italy; [14]Politecnico Milano, Piazza Leonardo da Vinci, 32 - 20133 Milano, Italy; [15]Dept. of Physics and Astronomy University of Padua, vicolo Osservatorio 3, 35122, Padova, Italy; [16]CEA Saclay, DSM/IRFU/SAp, 91191 Gif sur Yvette, France; [17] IAAT University of Tuebingen, Sand 1 - 72076 Tuebingen, Germany; [18]INPE, Avenida dos Astronautas 1.758, Jd. da Granja - 12227-010 São José dos Campos, Brazil; [19]National Space Institute, Technical University of Denmark, Elektrovej Bld 327, 2800 Kgs Lyngby, Denmark; [20] DAM and ICC-UB, University of Barcelona, Martì i Franques 1, 08028 Barcelona, Spain; [21]Arcetri Observatory, INAF, Largo Enrico Fermi 5 - I-50125 Firenze, Italy; [22]Cagliari University, Strada provinciale per Sestu, KM 1 - 09042 Monserrato, Italy; [23]Astronomical Institute of the Academy of Sciences of the Czech Republic, Fricova 298, CZ-251 65 Ondrejov, Czech Republic; [24]Cambridge University, Trinity Lane - CB2 1TN Cambridge, United Kingdom; [25]DPNC, Geneve University, Quai Ernest-Ansermet 30 - 1205 Geneva, Switzerland; [26]Laboratoire d'Astrophysique de Bordeaux, rue de l'Observatoire - BP 89 - 33270 Floirac Cedex, France; [27]INAF-OA Roma, Via Frascati, 33 - 00040 Monte Porzio Catone,

Italy; [28]MIT, 77 Massachusetts Avenue - MA 02139 Cambridge, Unitd States; [29]MSSL, Holmbury St Mary - RH5 6NT Dorking, Surrey, United Kingdom; [30]McGill University, 845 Sherbrooke Street West - QC H3A 0G4 Montréal, United States; [31]INAF-OA Capodimonte, Salita Moiariello, 16 - 80131 Napoli, Italy; [32]Ferrara University, Via Saragat 1 - 44122 Ferrara, Italy; [33]Florida University, - FL 32611 Gainesville, United States; [34]ISSI Bern, Hallerstrasse 6 - 3012 Bern, Switzerland; [35]Leicester University, University Road - LE1 7RH Leicester, United Kingdom; [36]Universities Space Research Association, 6767 Old Madison Pike, Suite 450, Huntsville, Alabama 35806, USA; [37] Monash Centre for Astrophysics, School of Physics and School of Mathematical Sciences, Monash University, Clayton VIC 3800, Australia; [38]Johns Hopkins University, 3400 North Charles Street - Baltimore, United States; [39] University of Tasmania, Private Bag 37, Hobart, TAS 7001, Australia; [40]MPA Garching, Karl-Schwarzschild-Str. 1 - 85741 Garching, Germany; [41]Radboud University, Heyendaalseweg 135 - 6500 GL Nijmegen, The Netherlands; [42]Clemson University, Clemson, SC 29631, United States; [43]Open University, Walton Hall - MK7 6AA Milton Keynes, United Kingdom; [44]Astrophysics Office, ZP12, NASA/Marshall Space Flight Center, Huntsville, AL 35812; [45]Department of Physics, Division of Geophysics and Astronomy, P.O. Box 48, FI-00014 University of Helsinki, Finland; [46]Durham University, Stockton Rd - DH1 3UP Durham, United Kingdom; [47]ICRA, Piazza della Repubblica, 10 - 65122 Pescara, Italy; [48]University of Iowa, 138 Iowa Memorial Un - IA 52242 Iowa City, United States; [49]Michigan state University, 567 Wilson Rd, - MI 48824 East Lansing, United States; [50]Copernicus Astronomical Center, Bartycka 18 - Warsaw, Poland; [5151]Science and Technology Office, ZP12, NASA/Marshall Space Flight Center, Huntsville, AL 35812; [52]Cornell University, Space Building - NY 14853 Ithaca, United States; [53] University of Illinois, Physics Department, 1110 W. Green St., Urbana, IL 61801, United States; [54]Dipartimento di Fisica, Università degli Studi di Milano, Via Celoria 16, 20133 Milano, Italy; [55]University of Trieste, Via Alfonso Valerio, 32 - 34128 Trieste, Italy; [56]Pavia University, Corso Strada Nuova, 65 - 27100 Pavia, Italy; [57]INAF IFC, Via Ugo La Malfa, 153 - 90146 Palermo, Italy; [58]University of California, Santa Cruz, 1156 High St. CA 95064 Santa Cruz, United States; [59]University of Rome III, Via della Vasca Navale, 84 - 00146 Roma, Italy; [60]University of Melbourne, Swanston Street - VIC 3052 Parkville, Australia; [61]Kapteyn Astronomical Institute, University of Groningen, P.O. Box 800, 9700 AV Groningen, The Netherlands; [62]University of Maryland, Department of Astronomy College Park, MD 20742-2421, United States; [63]National Institute of Aerospace Technology (INTA), Carretera de Ajalvir km. 4 - 28850 Torrejón de Ardoz, Spain; [64]University of Alberta, 85 Avenue 116 St NW - AB T6G 2R3 Edmonton, Canada; [65]Observatoire Astronomique de Strasbourg, 11 rue de l'Université - 67000 Strasbourg, France; [66]Université Paris Diderot 5 rue Thomas-Mann 75205 Paris cedex 13, France; [67]INAF-OA Torino, Via Osservatorio, 20 - 10025 Pino Torinese, Italy; [68]Space Research Centre, Warsaw, Bartycka 18A - Warszawa, Poland; [69]Space Telescope Institute, 3700 San Martin Drive - MD 21218 Baltimore, United States; [70]University of Arizona, Department of Astronomy, 933 N. Cherry Ave, Tucson, AZ 85721, United States; [71]Raman Research Institute, C. V. Raman Avenue - 560 080 Sadashivanagar, India; [72]Czech Technical University in Prague, Zikova 1903/4, CZ-166 36 Praha 6, Czech Republic; [73]Department of Physics, Division of Astronomy, FI-90014 University of Oulu, Finland; [76]Armagh Observatory, College Hill - BT61 9DG Armagh, United Kingdom; [77]INFN, Trieste, Via A. Valerio 2 - I-34127 Trieste, Italy; [78]NRL, 4555 Overlook Ave. SW Washington, DC 20375-5352, United States; [80]University of Washington, 3910 15th Ave. NE - Seattle, United States; [81]Physics Department, University of Crete, GR-710 03 Heraklion, Greece; [82]Istituto de Astrofisica de Canarias, Vía Láctea s/n, La Laguna, E-38205, Tenerife, Spain; [83]Leiden Observatory, Niels Bohrweg 2 - NL-2333 CA Leiden, The Netherlands; [84]KTH Royal Institute of Technology, Valhallavägen 79 - 100 44 Stockholm, Sweden; [85]INAF-IASF-Milano, Via E. Bassini 15 - I-20133 Milano, Italy; [86]Silesian University in Opava, Na Rybníčku 626/1 - 746 01 Opava, Czech Republic; [87]University of Erlangen-Nuremberg, Schlossplatz 4 - 91054 Erlangen, Germany; [88] ExtreMe Matter Institute EMMI, GSI Helmholtzzentrum für Schwerionenforschung GmbH, 64291 Darmstadt, and Institut für Kernphysik, Technische Universität, 64289 Darmstadt, Germany; [89]Leibniz-Institut fuer Astrophysik Potsdam, An der Sternwarte 16 - 14482 Potsdam, Germany; [90]National University of Ireland Galway, University Road, Galway, Ireland; [91]University of Manchester, Booth Street West - M15 6PB Manchester, United Kingdom; [92]Aristotle University of Thessaloniki, Greece; [93]Goddard Space Flight Center, 8800 Greenbelt Rd. - Md., 20771 Greenbelt, United States; [94]ISAS, 3-1-1 Yoshinodai, Chuo-ku, Sagamihara - 252-5210 Kanagawa, Japan; [95]University of Alicante, Carretera San Vicente del Raspeig - 03690 Sant Vicent del Raspeig, Spain; [96]ICREA - Institució Catalana de Recerca i Estudis Avançats, Passeig Lluís Companys, 23 - 08010 Barcelona, Spain; [97]Physics Institute of the Academy of Sciences of the Czech Republic, Na Slovance 1999/2, CZ-182 21 Praha 8, Czech Republic; [98]University of Warwick, Gibbet Hill Road - CV4 7AL Coventry, United Kingdom; [99]INAF-OA Padova, Vicolo Osservatorio 5, Padova, Italy; [100]University of Nevada, Las Vegas, NV 89012, United States; [101]University of Rome Tor Vergata, Via della Ricerca Scientifica 1 - 00133 Rome, Italy; [102]INAF-IRA-Bologna, Via P. Gobetti 101- 40129 Bologna, Italy; [103]University of Bologna, Dept. of Physics and INFN section of Bologna, V.le Berti Pichat, 6/2, 40127, Bologna, Italy; [104]Departamento de Astrofisica, Universidad de La Laguna, La Laguna, E-38206, Santa Cruz de Tenerife,

Spain; [105]APC, AstroParticule & Cosmologie, UMR 7164 CNRS/N2P3, Université Paris Diderot, CEA/Irfu, Observatoire de Paris, Sorbonne Paris Cite 10 rue Alice Domon et Leonie Duquet, 75205 Paris Cedex 13, France; [106]INAF- Osservatorio Astronomico di Palermo, Piazza del Parlamento 1, 90134 Palermo; [107]School of Physics and Astronomy, University of Southampton, Southampton, SO17 1BJ, UK; [108]Kepler Institute of Astronomy, University of Zielona Gòra, Lubuska 2, 65-265, Zielona Gòra, Poland; [109]University of California, San Diego, Mail Code 0424, La Jolla, CA 92093 United States; [110]Dept. of Astronomy, Univ. of Wisconsin, 475 N. Charter Str., Madison WI 53706; [111]Wayne State University, Department of Physics & Astronomy, 666 W. Hancock St, Detroit, MI 48201, USA; [112]Centro de Astrobiologia (CSIC-INTA), P.O. Box 78, E-28691, Villanueva de la Cañada, Madrid, Spain; [113]Foundation for Research and Technology - Hellas, GR-711 10 Heraklion, Greece; [114]CNES, 18 Avenue Edouard Belin, 31400 Toulouse, France.

[*] marco.feroci@inaf.it

## ABSTRACT

The LOFT mission concept is one of four candidates selected by ESA for the M3 launch opportunity as Medium Size missions of the Cosmic Vision programme. The launch window is currently planned for between 2022 and 2024. LOFT is designed to exploit the diagnostics of rapid X-ray flux and spectral variability that directly probe the motion of matter down to distances very close to black holes and neutron stars, as well as the physical state of ultradense matter. These primary science goals will be addressed by a payload composed of a Large Area Detector (LAD) and a Wide Field Monitor (WFM). The LAD is a collimated (<1 degree field of view) experiment operating in the energy range 2-50 keV, with a 10 $m^2$ peak effective area and an energy resolution of 260 eV at 6 keV. The WFM will operate in the same energy range as the LAD, enabling simultaneous monitoring of a few-steradian wide field of view, with an angular resolution of <5 arcmin. The LAD and WFM experiments will allow us to investigate variability from submillisecond QPO's to year-long transient outbursts. In this paper we report the current status of the project.

## 1. INTRODUCTION

The LOFT mission concept (Large Observatory For x-ray Timing, Feroci et al. 2011) was submitted on December 2010 in response to the M3 call issued by the European Space Agency (ESA) within the framework of the Cosmic Vision 2015-2025 programme. LOFT is designed to observe the rapid spectral and flux variability of X-rays emitted from regions close to the surface of neutron stars and the event horizons of black holes. The proposed measurements are efficient diagnostics of the behavior of matter in the presence of strong gravitational fields, where the effects predicted by General Relativity are largest, and the physics of matter at densities in excess of that in atomic nuclei, determining its equation of state and composition. This research addresses science theme 3 proposed in the ESA Cosmic Vision: "What are the fundamental physical laws of the Universe". The LOFT mission was selected as one of 4 mission candidates for a single launch opportunity in the time frame 2022-2024. System aspects are currently being studied by ESA and its industrial contractors. The scientific payload is being studied by a consortium of European scientific institutes, including teams from the Czech Republic, Denmark, Finland, France, Germany, Italy, the Netherlands, Poland, Spain, Switzerland, and the United Kingdom, with support from international partners in Brazil, Japan and the United States. An even wider science support community (Figure 1 shows the current world-wide geographic distribution of the "LOFT community") is contributing by providing scientific inputs to help focus and refine the science case and the scientific requirements. ESA currently plans to select one sole M3 mission candidate by the end of 2013.

The scientific payload of the LOFT mission includes two experiments: the Large Area Detector (LAD, Zane et al. 2012) and the Wide Field Monitor (WFM, Brandt et al. 2012). The key feature of the LAD is its very large area, ~20 times larger than any predecessor ever flown, combined with "CCD-class" energy resolution. The 1-deg collimated field of view LAD will be able to access ≥50% of the sky at any time, to observe the most interesting Galactic and extragalactic sources in their most interesting states. To guarantee this, LOFT is equipped with the WFM, which will monitor more than half of the LAD-accessible sky (approximately 1/3 of the whole sky) simultaneously at any time. The WFM operates in the same energy range as the LAD, providing information about source status (flux variability and energy spectrum), as well as arc-minute positioning. With such wide angle sky monitoring, the WFM will also provide long-term histories of the target sources, serving to facilitate both the LAD observations and a series of wider science goals.

In this paper we provide an overview of the LOFT mission, including the science drivers and goals, a short description of the payload and the mission profile. More detailed descriptions of the LAD and WFM instruments may be found in Zane et al. 2012 and Brandt et al. 2012, respectively.

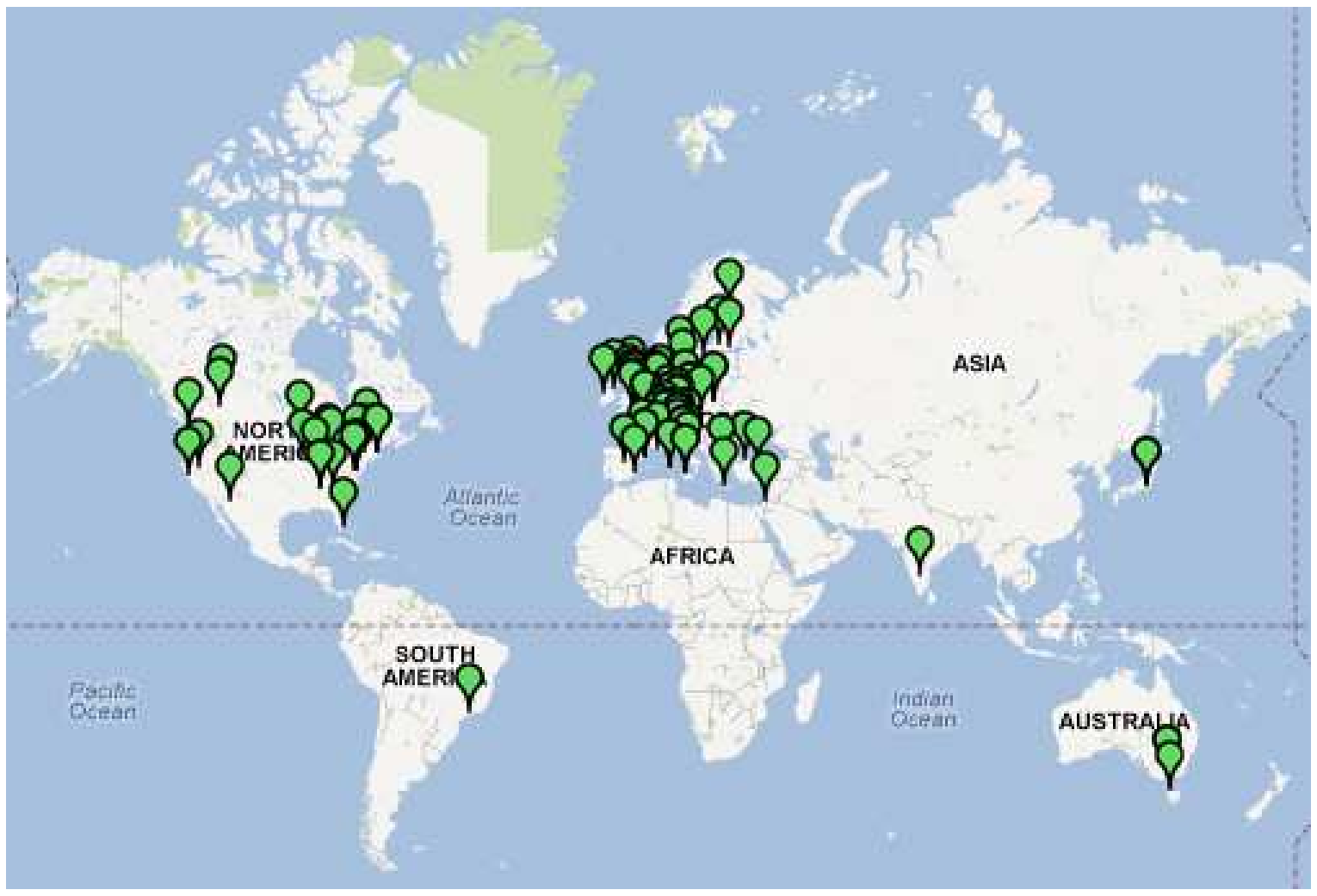

Figure 1. The geographic distribution of the LOFT Community.

## 2. SCIENCE OBJECTIVES

The main science drivers of the LOFT mission are the study of matter in ultradense environments and under strong gravitational fields. These two major themes are the main science objectives, driving the design of the mission. However, the high performance of LOFT makes it an ideal observatory for a wide range of celestial objects. In the following sections we summarize very briefly the main science objectives for these three “categories”: “equation of state”, “strong gravity”, and “observatory science”, listing the quantitative top-level science goals identified by the LOFT Science Study Team. A more extended discussion on these themes will be reported in forthcoming papers in preparation by the LOFT science working groups.

### 2.1 Equation of state of ultradense matter

One of the most important open questions in high energy astrophysics and nuclear physics is the relation between density and pressure of matter at densities larger than that of atomic nuclei, where quantum chromodynamics (QCD) and quark physics are poorly constrained. This high-density/low-temperature region of the QCD phase diagram is inaccessible to high-energy terrestrial experiments and can only be probed in neutron stars (NS). Many equation of state (EOS) models have been proposed over the years, which predict how the mass M and the radius R of NS are related. Recent accurate measurements of high NS masses in radio pulsars have provided significant constraints, but the issue of determining the equation of state – which requires both M and R - is still wide open.

While more precision mass estimates are expected from highly accurate radio measurements, determination of NS radius or simultaneous measurements of M and R in the same object, are most likely to come from X-ray observations. A 10 $m^2$ -class observatory offers different independent tools to achieve this goal. Of these, modeling of pulse profiles is probably the most promising. The shape of the coherent pulsed signal in accreting millisecond pulsars and during thermonuclear bursts encodes effects related to M/R, as the photons propagate through curved space-time. Modeling of these distortions (Doppler boosting, time dilation, gravitational light bending and frame dragging), enabled by high statistics, will provide accurate measurements of M and R. The large collecting area will also enable use of complementary techniques exploiting continuum spectral modeling in photospheric radius expansion bursts, global seismic oscillations during intermediate flares of magnetars (detected serendipitously from outside the field of view), and a more complete NS spin distribution.

The key LOFT requirements with respect to determining the dense matter EOS have been quantified as 3 top-level science goals:

*EOS1* Constrain the Equation of State of supranuclear-density matter by the measurement, using three complementary types of pulsations, of mass and radius of at least 4 NS with an instrumental accuracy of 4% in mass and 3% in radius.

*EOS2* Provide an independent constraint on the EoS by filling out the accreting NS spin distribution through discovering coherent pulsations down to an amplitude of about 0.4% (2%) *rms* for a 100 mCrab (10 mCrab) source in a time interval of 100 s, and oscillations during type I bursts down to typical amplitudes of 1% (2.5%) *rms* in the burst tail (rise) among 35 NS covering a range of luminosities, inclinations and binary orbital phases.

*EOS3* Probe the interior structure of isolated NS by observing seismic oscillations in Soft Gamma-ray Repeater intermediate flares with flux ~1000 Crab through high energy photons (> 20 keV).

## 2.2 Behavior of matter under strong gravity

The other main science driver for LOFT is the study of the behavior of matter in the presence of strong gravitational fields (strong-field gravity, SFG). General Relativity (GR) has been probed and proven to a high degree of accuracy only in the weak-field regime, for gravitational radii $r_g \sim 10^5$-$10^6$. X-rays emitted from the innermost regions (a few $r_g$) around compact objects (neutron stars and black holes) originate from matter experiencing a strong gravitational potential and provide the best tools to explore the physics in such extreme conditions, testing the predictions of GR where they are expected to produce macroscopic effects. LOFT will be able to approach this science goal from different angles, using both its large area and its high spectral resolution. Black holes and neutron stars display quasi-periodic oscillations (QPOs) in their X-ray flux arising from the millisecond dynamical timescales of the inner accretion flows. The interpretation of such high frequency QPOs necessarily involves fundamental frequencies of the motion of matter orbiting in disk regions dominated by the gravitational field. Examples of such interpretations are the competing models attributing high frequency QPOs to relativistic radial and vertical epicyclic frequencies versus those predicting them to arise from relativistic nodal and periastron precession. The LOFT measurements will discriminate between such models and directly access so far untested GR effects, such as frame-dragging, strong-field periastron precession, and the existence of an innermost stable orbit around black holes.

LOFT will open a new era in the field of X-ray timing, providing access to information that is qualitatively new, due to its capability to measure dynamical timescale phenomena on their coherence time: previously we have been able to study only time-averaged behavior. One beautiful example serves to illustrate LOFT's technical capabilities. By combining large area with good spectral energy resolution, low frequency (e.g. ~30 Hz) QPOs will be detected in neutron star binaries at such high statistical accuracy to allow true phase-resolved spectroscopy, as for a coherent signal. The observation of the variable Fe K line profile at different phases will permit detection of the expected features of Lense-Thirring precession of the inner disk at $\sim r_g$, providing a measurement of the inclination of the varying ring.

The variability of the Fe K line profile is also a tool to measure mass and spin of black-holes. This is possible using both Galactic and extragalactic black holes, but is best done using Active Galactic Nuclei, where the longer dynamical timescales compensate for their dimmer flux, providing better counting statistics per dynamical timescale. Reverberation mapping measurements in a few tens of bright AGN are expected to provide significant constraints on the mass and spin of their supermassive black holes.

The key LOFT requirements with respect to SFG have been quantified as 5 top-level science goals:

*SFG1* Detect strong-field GR effects by measuring epicyclic motion in high frequency QPOs from at least 3 black hole X-ray binaries.

*SFG2* Detect disk precession due to relativistic frame dragging with the Fe line variations in low frequency QPOs for 10 NSs and 5 BHs.

*SFG3* Detect kHz QPOs at their coherence time, measure the waveforms and quantify the distortions due to strong-field GR for at least 10 NSs covering different inclinations and luminosities.

*SFG4* Measure the Fe-line profile and carry out reverberation mapping of 5 BHs in binaries to provide BH spins to an accuracy of 5% of the maximum spin (a/M=1), constraining fundamental properties of stellar mass black holes and of accretion flows in strong field gravity.

*SFG5* Measure the Fe-line profile of 30 AGNs, and carry out reverberation mapping of the 8 AGNs most suitable for the latter purpose, to provide BH spins to an accuracy of 20% of the maximum spin (10% for fast spins) and measure their masses with 30% accuracy, constraining fundamental properties of supermassive black holes and of accretion flows in strong field gravity.

The LOFT Science Study Team has translated the top-level goals EOS 1-3 and SFG 1-5 into a sequence of realistic observations that will enable us to meet the science requirement. Table 1 shows the list of targets, by class and category, identifying the number of targets, whether the observation can be planned in advance or must be a Target of Opportunity (ToO), the anticipated number of pointings and the total observing time necessary to reach the required science goal. ToOs can of course only be predicted on probabilistic grounds. On the basis of the known statistics, the plan given in Table 1 ensures a 99% chance of detecting high frequency QPOs in at least one Black-Hole transient, and a 94% of detecting them in two.

Table 1. The preliminary breakdown of the LOFT observing program necessary to satisfy the Core Science requirements. A nominal 4-year mission lifetime is assumed as a baseline. The total amount of observing time required by the Core Science is 21 Ms, corresponding to 40% of the total available observing time. A significant fraction of this time (in addition to the remaining 60%) will be part of the Guest Observer Program.

| Source Type | ToO | Number of Sources | Number of Pointings | Total Time (ks) | Science Goal |
|---|---|---|---|---|---|
| BH transient outburst | Yes | 4 | 800 | 2400 | SFG 1,2, 4 |
| Persistent BH | No | 2 | 400 | 1600 | SFG 1, 2, 4 |
| AGN | No | 30 | 50 | 8000 | SFG 5 |
| msec pulsar outburst | Yes | 3 | 250 | 1000 | EOS 1, SFG 2, 3 |
| NS transient bright outburst | Yes | 3 | 250 | 1800 | EOS 1, 2, SFG 3 |
| Persistent bright NS | No | 12 | 350 | 4800 | EOS 1, 2, SFG 2, 3 |
| NS transient weak outbursts | Yes | 6 | 6 | 120 | EOS 2 |
| Persistent weak NS | No | 14 | 14 | 280 | EOS 2 |
| Bursters | Yes | 10 | 40 | 1000 | EOS 2 |

### 2.3 LOFT as a general observatory

The EOS and SFG areas are the primary LOFT science drivers, meaning that they set the tightest requirements on the main instrumental properties, such as the effective area or spectral resolution. Clearly, an instrument like the LAD – with CCD-class energy resolution over a wide band, combined with 10 $m^2$ effective area – or the WFM – with soft X-ray bandpass, ~300 eV energy resolution and arcmin imaging simultaneously over ~3-4 steradians – will dramatically change the observational scenario for a wide range of both Galactic and extragalactic sources. Time variability and spectroscopy will become accessible with the LAD on as yet unexplored timescales, while the WFM will provide spectral information on unpredictable flares and bursts from Galactic sources as well as cosmic gamma-ray bursts. An overview of the entire potential discovery space of LOFT as a general observatory is impossible here, but below we give a few examples.

X-ray binaries are amongst the most "obvious" targets for LOFT, with a outstanding range of science questions that might be addressed. Periodicities and period evolution will be detected at unprecedentedly low amplitude levels. Transient periodicities (such as the intermittent pulsations detected in Aql X-1) will be searched for and most likely detected from several candidates. The enormous counting statistics offered by the LAD will allow detailed studies of the physics and geometry of accretion onto magnetized objects. The WFM will provide long-term histories for these studies but will also be able to trigger early LAD observations of outbursts of jet sources. With the advent of high time resolution optical and infrared instruments on large telescopes, multi-wavelength studies will provide information on jet speeds even in those cases in which the jet is not spatially resolved.

Whilst the WFM will offer long-term flux and spectral monitoring of the bright AGNs, LAD pointed observations of the highly variable blazars will be able to provide measurements of their flux and spectral variability down to timescales of seconds, as yet unexplored in X-rays. Meanwhile continuum spectra, as well as Fe K features, in ~mCrab-AGNs will be studied with high accuracy up to 30 keV, with good sensitivity to the Compton reflection component.

The properties of the WFM – energy range, energy resolution and field of view – make it ideal for the detection, localization and measurement of the prompt emission in gamma-ray bursts. More than a hundred events per year are anticipated in the WFM field of view. The sensitivity to soft X-rays, combined with good spectral resolution, will prove crucial to confirm (or refute) the transient absorption edge features in the prompt emission (which if detected would provide independent distance estimates). The high-significance study of the time-dependent continuum energy spectrum will not only enable us to put important constraints on the physics of the prompt GRB emission but also carry information on the circumburst environment and thus on the progenitors. Last but not the least, the soft X-ray bandpass is essential to permit studies of soft gamma-ray bursts, significantly increasing the detection rate of high-redshift events.

The LOFT mission and instrumentation are conceived so that the well-focused breakthrough science expected in the fields of the physics of ultradense matter and strong gravity can be achieved with high flexibility and versatility that exploit to best advantage the large potential of a unique combination of effective area and energy resolution, supported by a high performance wide field monitor. Table 2 summarizes the basic mission requirements necessary to satisfy the core science requirements. In the next sections we provide a description of the payload instruments and mission characteristics.

Table 2. The LOFT mission requirements.

| Parameter | Value |
|---|---|
| Net observing time for core science | 20.2 Ms |
| Additional open observing time | >20 Ms |
| Mission duration | 4 years |
| Accessible sky fraction (LAD) | >50% |
| Orbit | LEO, <600 km, <5° |
| ToO (following alert of SOC) | <12 hrs working hours<br><24 hrs otherwise |
| Slews per orbit | 2 |
| Data transfer | 6.7 Gbit/orbit |

## 3. SCIENTIFIC PAYLOAD

The science objectives briefly summarized in the previous section and quantified in the scientific requirements as EOS and SFG will be addressed by a scientific payload composed of the LAD and WFM instruments. Figure 2 shows a pictorial view of the LOFT satellite, in the configuration currently being studied by the LOFT Consortium. The six LAD panels are deployed, and the five units of the WFM are sitting on the optical bench. Compared to the original LOFT proposal (Feroci et al. 2011), the first ~year of study of the mission, including the focused assessment by the Concurrent Design Facility of ESA, has not changed the general configuration of the mission and payload. The only exception is the WFM design, which has evolved to cover a much larger field of view, going from 2 to 5 offset units.

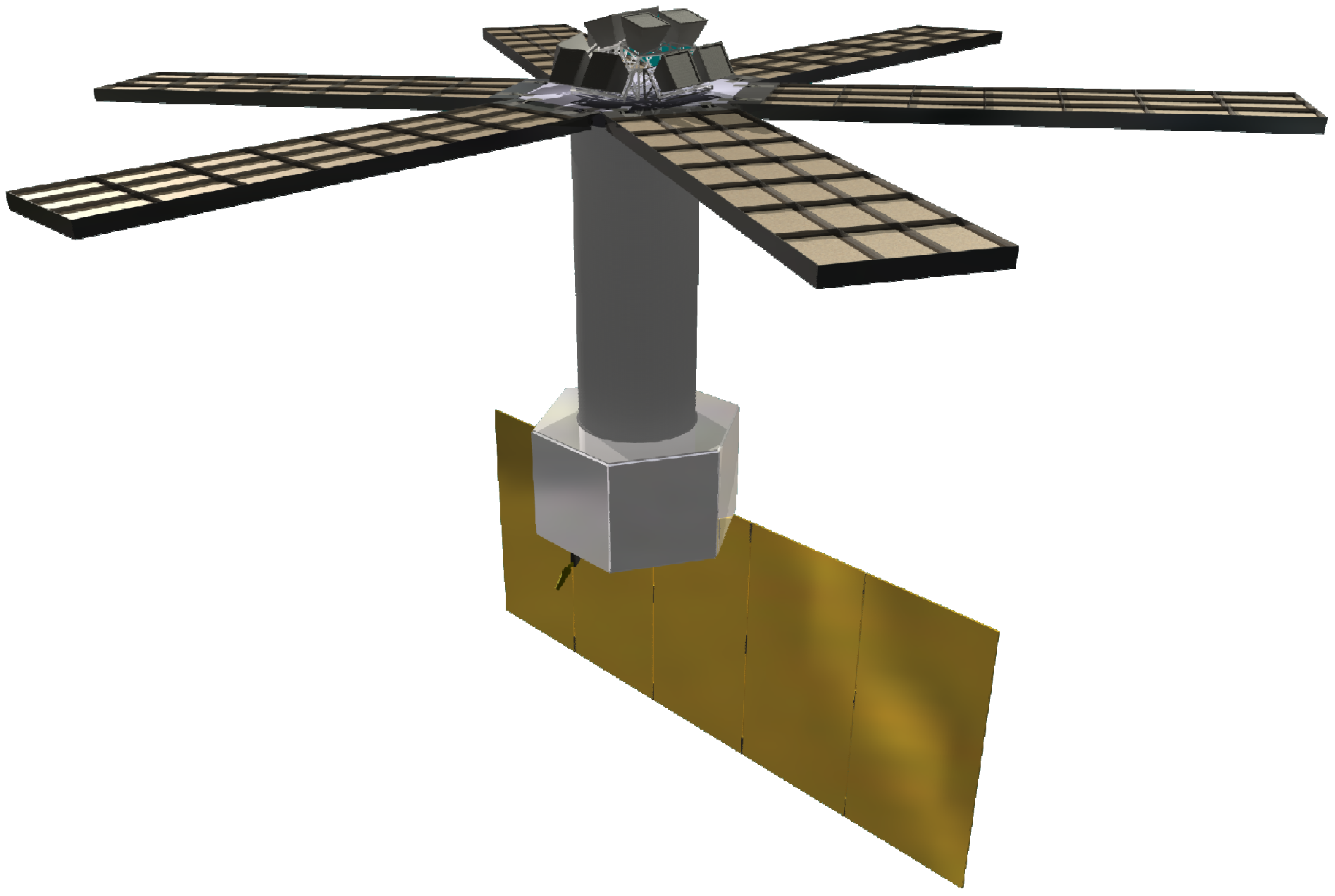

Figure 2. A pictorial view of the LOFT mission currently being studied by the LOFT Consortium, showing the deployed LAD panels, including the detector modules, as well as the 10 cameras comprising the 5 units of the WFM. The structure of 21 Modules in each Panel is shown, although the individual detector tiles are not visible. A structural tower supports the optical bench, with the service module supporting the solar panel array.

### 3.1 The Large Area Detector

The LAD is the prime instrument onboard LOFT. Driving its performance are the effective area and the energy resolution. The LAD has a geometric area as large as 15 $m^2$, achieving a peak effective area of 10 $m^2$ while offering an energy resolution (FWHM) better than 260 eV @ 6 keV. The large area of the LAD is made affordable within the context of the resource budgets of a medium-class mission by its two technology drivers - the large area Silicon drift detectors (SDDs) and the capillary plate collimators – enabling a very high effective area per unit mass/volume/power. The overall budgets of the LOFT are not so different from those of the Rossi X-ray Timing Explorer (Jahoda et al. 2006), but the effective area of the LAD is approximately 20 times larger than that of the Proportional Counter Array.

The large-area SDDs were originally developed (Vacchi et al. 1991, Rashevski et al. 2002) for particle tracking in the Inner Tracking System of the ALICE experiment at the Large Hadron Collider at CERN: 1.5 $m^2$ of SDDs in ALICE have been operating successfully since 2008. In that context, they are monolithic Silicon detectors with ~50 $cm^2$ effective area, read-out by two series of anodes. In the LOFT/LAD application, the SDDs are 450 μm thick, have an active area of 76 $cm^2$ each, read-out by two rows of anodes with a pitch of 970 μm. The drift channel is 35 mm long, resulting in an area of 0.3 $cm^2$ per read-out channel. The working principle of such detectors, described in detail in Zampa et al. 2011, is as follows: an X-ray photon is photo-electrically absorbed in the Silicon bulk, locally generating a charge cloud. An electric field, sustained by a voltage drop of ~1300V from the median plane of the Silicon tile to each of the edges hosting the anode rows, makes the electrons to drift from the absorption point to the anodes. While drifting, the diffusion causes the charge cloud to widen, up to ~1 mm for photons absorbed near the middle plane of the detectors, 35 mm away from the anodes. This implies that the total charge may be collected by either 1 or 2 anodes, depending on the impact point of the photon (both in the drift direction as well as in the anode direction). In the LAD configuration, approximately 45% of the events are collected by a single anode, while 55% are collected by two anodes: we classify them as singles or doubles, according to their multiplicity. As the signal-to-noise is related to the performance of each read-out anode, single events confront a √2 smaller noise than double events. For this reason, a fraction of 45% of the LAD area offers an energy resolution higher than the total events. In the LAD read-out architecture, this information is identified and preserved, enabling spectral studies with higher performance at the cost of ~half the effective area. The time tagging of the individual event is done at the time of its discrimination at the anode preamplifier. This implies an uncertainty related to the drift time, depending on the impact point. The maximum uncertainty is ~7 μs, for events that drift along the whole channel.

The other key technology is the capillary-plate collimator. This is actually not new, as similar devices have already been used by the MEDA and GSPC onboard the EXOSAT mission and it are currently baselined for the MIXS experiment onboard the ESA BepiColombo mission (Fraser et al. 2010). In the LOFT application, they adopt the manufacturing technology of the micro-channel plates: a thin disk of lead glass with millions of micro-pores. The Pb content in the glass offers enough stopping power to X-rays, making it suitable as collimator for soft X-rays. The LOFT LAD configuration envisages square pores with 100 μm opening and 20 μm walls, compliant with the requirement of <1% transparency at 30 keV. The plate thickness is 6 mm, reaching the aspect ratio of 60:1 corresponding to a field of view of 1 degree FWHM. The open area ratio is 70% and the size of each collimator tile is such to cover a whole SDD, approximately 70 mm x 110 mm. While offering a very compact and light collimator system, the stopping power of capillary plates becomes relatively poor to hard X-rays (above ~30-40 keV). For this reason, the anticipated background of the LAD is dominated by the counts induced by hard X-ray photons in the Cosmic X-ray Background, as well as those from the Earth albedo. A detailed analysis of the LAD background sources, together with Monte Carlo estimates, is reported in Campana et al. (2012).

Both the detector and the collimator components in the LAD are individual tiles, ~80 $cm^2$ in area. The LAD is then intrinsically modular (and thus redundant). As it shown in Figure 2, the overall instrument is organized in 6 Detector Panels, each one hosting 21 Detector Modules, in turn composed of 16 SDDs and collimator tiles. Each Module is equipped with its own Module Back End Electronics (MBEE) interfacing the Front End Electronics (FEE) of the 16 detectors. The MBEE is in charge of providing regulated power and digital commands, as well as handling the data I/O. The 21 MBEEs of each panel are interfaced by a Panel Back End Electronics (PBEE), in turn connected with the single LAD Data Handling Unit (DHU). A detailed description of the LAD digital electronics and its architecture is given in Suchy et al. (2012).

The basic unit of the LAD is the Module. This is organized in a mechanical frame providing assembly and alignment interface to the 16 detectors and to the relevant MBEE. Each detector is composed of a SDD equipped with its own FEE in a sandwich architecture. The other main component of a Module is the collimator frame. Similar to the detectors, each of the 16 collimator tiles is interfaced and aligned to a common mechanical support, forming a single overall grid-like structure. The detector and collimator trays are then integrated and aligned, forming a complete Module. It is worth noticing that the field of view of the LAD instrument is given by the collimators (not the SDDs) and for this reason these are the systems requiring a careful alignment. It is planned to achieve this goal through an isostatic mount of each Module in the relevant Panel structure, a technique able to provide arcsecond alignment accuracies. Additional details on the LAD mechanical layout may be found in Zane et al. (2012).

The overall LAD is thus composed of 126 Modules, for a total surface of 18 $m^2$, including 15 $m^2$ of Silicon detectors. Each of the Modules is electrically independent, although commanded through a common PBEE with the other Modules in the same panel. The total effective area of the LAD as a function of energy is shown in Figure 3, which accounts for all the factors reducing exposed detector area, from the blocking of the collimator to the electrodes on the surface of the Silicon detector. At launch, the 6 LAD panels are stowed around the support tower to fit the rocket fairing.

The use of a highly segmented detector means that although the LAD will collect ~240000 cts/s while observing the Crab, each read-out channel (which has an area of 0.3 $cm^2$) will only detect a few counts/s, making pile-up completely negligible. In principle, exploiting the same segmentation could also render dead-time completely negligible. In practice, this would require an independent handling of each channel in the read-out ASICs, increasing complexity. As a trade-off, we identified half of an SDD tile (114 channels) as the common read-out unit, driving the dead-time. With such a configuration, the dead-time during the observation of a source with a flux of 1 Crab is ~0.7%. Figure 4 is a flow diagram where count and data rates are shown at each step, from the detector, through the MBEE, the PBEE and the DHU, highlighting the advantages of the LAD segmentation.

In Table 3 we summarize the main scientific requirements of the LAD.

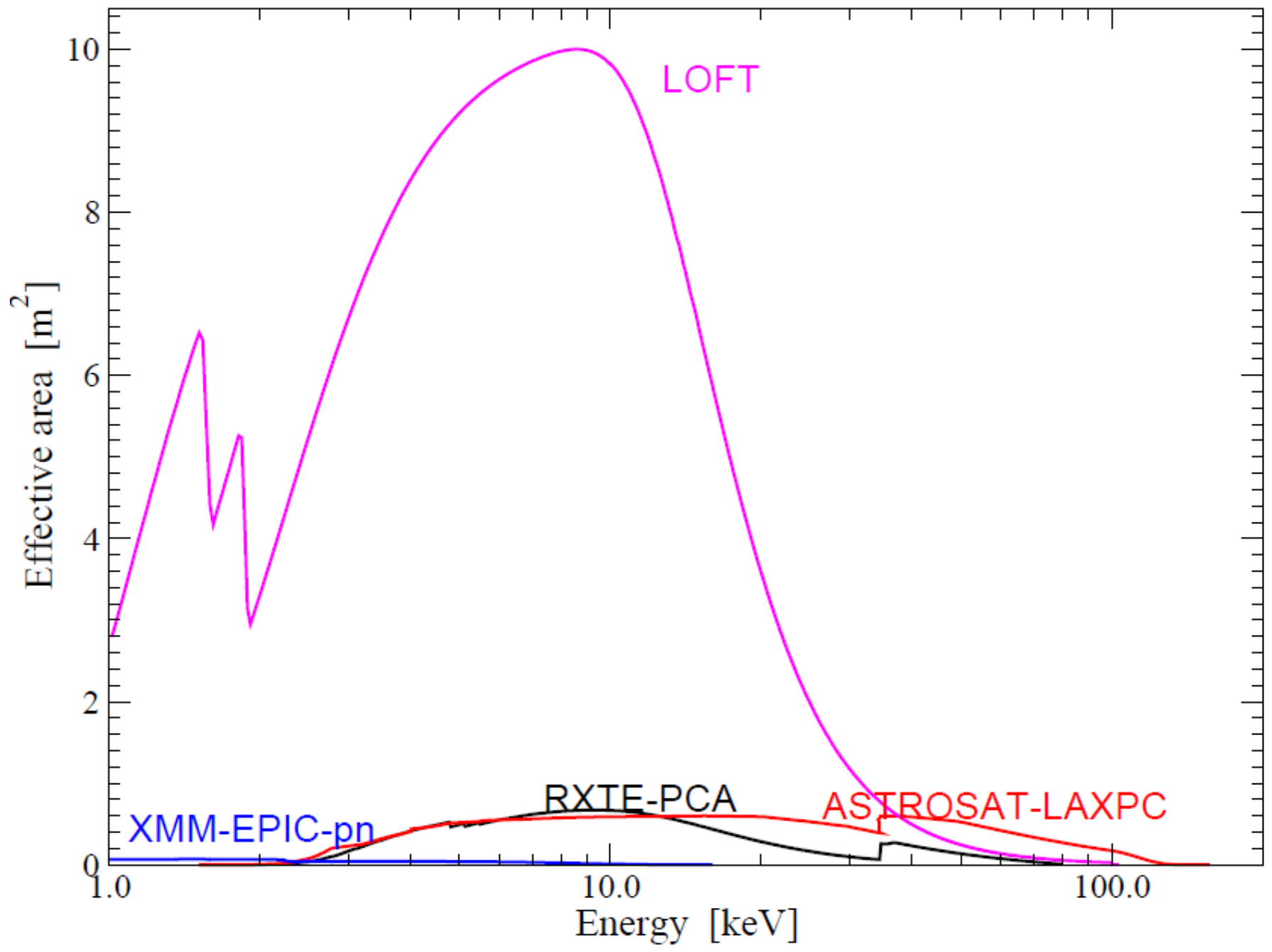

Figure 3. The effective area of the LAD as a function of energy, compared to previous and planned large-area X-ray missions. The response is shown down to 1 keV, although the baseline low energy discrimination threshold is currently set at 2 keV (studies are in progress to push the low energy response).

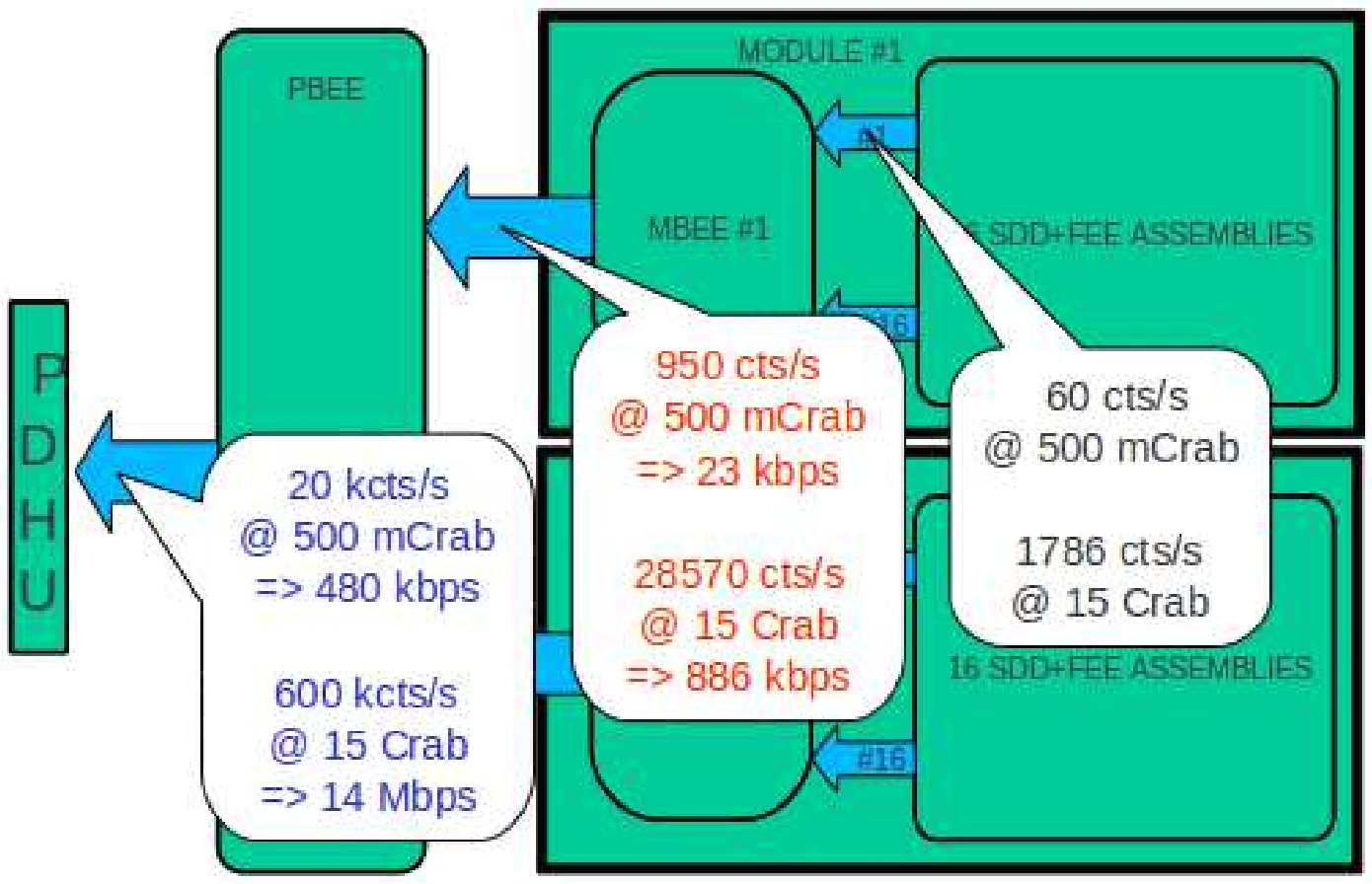

Figure 4. The LAD data flow, from the individual detectors (far right), through the MBEE, to the PBEE up to the DHU (far left), for two cases of source intensity: 0.5 and 15 Crab.

Table 3. The main scientific requirements of the LAD.

| Parameter | Value |
|---|---|
| Effective Area | 4 m$^2$ @ 2 keV<br>8 m$^2$ @ 5 keV<br>10 m$^2$ @ 8 keV<br>1 m$^2$ @ 30 keV |
| Energy Range | 2-30 keV primary<br>30-80 keV extended |
| Energy Resolution FWHM | 260 eV @ 6 keV<br>200 eV @ 6 keV (45% of area) |
| Collimated Field of View | 1 degree FWHM |
| Time Resolution | 10 micro-second |
| Absolute Time Accuracy | 1 micro-second |
| Dead Time | <1% @ 1 Crab |
| Background | <10 mCrab (<1% systematic) |
| Max Flux | 500 mCrab full event info<br>15 Crab binned mode |

### 3.2 The Wide Field Monitor

The WFM is based on the classical coded mask imaging technique. The specific LOFT design is an evolution of the design adopted in the SuperAGILE experiment (Feroci et al. 2007), with a noticeable improvement provided by the low energy threshold, better energy resolution and the (asymmetric) 2D imaging capabilities of the SDDs. In fact, as discussed in detail in Campana et al. (2011), the same SDDs used in the LAD can be used for imaging purposes in the WFM, by adopting a proper anode pitch (145 μm). With the same "1D" read-out electronics as the LAD, the WFM SDDs are able to localize the photon impact point with an accuracy as high as ~70 μm in the anode direction (by charge barycentering) and a ~3-8 mm, energy dependent, in the drift direction. The latter is achieved "for free" (that is, without any additional read-out) by measuring the width of the charge cloud reaching the anodes: the farer is the absorption point along the drift channel, the wider is the charge distribution at the anodes, as due to diffusion. A detailed discussion and analysis of the imaging properties of the SDDs for LOFT/WFM may be found in Evangelista et al. (2012). Indeed, the choice of a finer anode pitch is optimized to the best sampling of the charge distribution for the event position reconstruction. Combining these detectors with a proper, asymmetric coded mask at a distance of ~20 cm and a collimator blocking the diffuse X-ray background, the resulting camera is sensitive in the 2-50 keV[1], with an angular resolution of ~4 arcminute in the fine direction and ~5° in the coarse direction. For an optimal imaging of sources, two identical cameras, observing the same region of the sky but with the fine imaging direction rotated by 90°, form each individual WFM Unit. By combining the response of the two Cameras, a 2D angular resolution of 4'x4' is achieved, while keeping a strong redundancy in case of failure of one Camera (although with coarse resolution in one direction). The imaging properties of this system are extensively discussed in Donnarumma et al. (2012). The WFM detectors are the same as the LAD and as such they have similar intrinsic energy resolution. However, the choice of a finer pitch for imaging optimization implies the detection of the same charge with a larger number of anodes and therefore with a higher read-out noise. The resulting energy resolution is about 300 eV FWHM @ 6 keV (end of life, at the WFM operating temperature).

The main science requirement for the WFM experiment is to monitor and image the sky accessible to the LAD, to trigger its observations of the most interesting source states. In coded mask experiments, the sensitivity is driven by the aperture background and vignetting factors for large off-axis angles. Large regions of the sky are then better monitored by a set of smaller units. This is indeed the strategy adopted by the WFM, which covers simultaneously more than 50% of the sky accessible to the LAD with a set of 5 Units in off-set, each one composed of 2 Cameras, as shown in Figure 2. To optimize WFM sensitivity to weak sources, the open fraction of the coded masks has been chosen to be 25%. The resulting sky-projected WFM area is shown graphically in Figure 5.

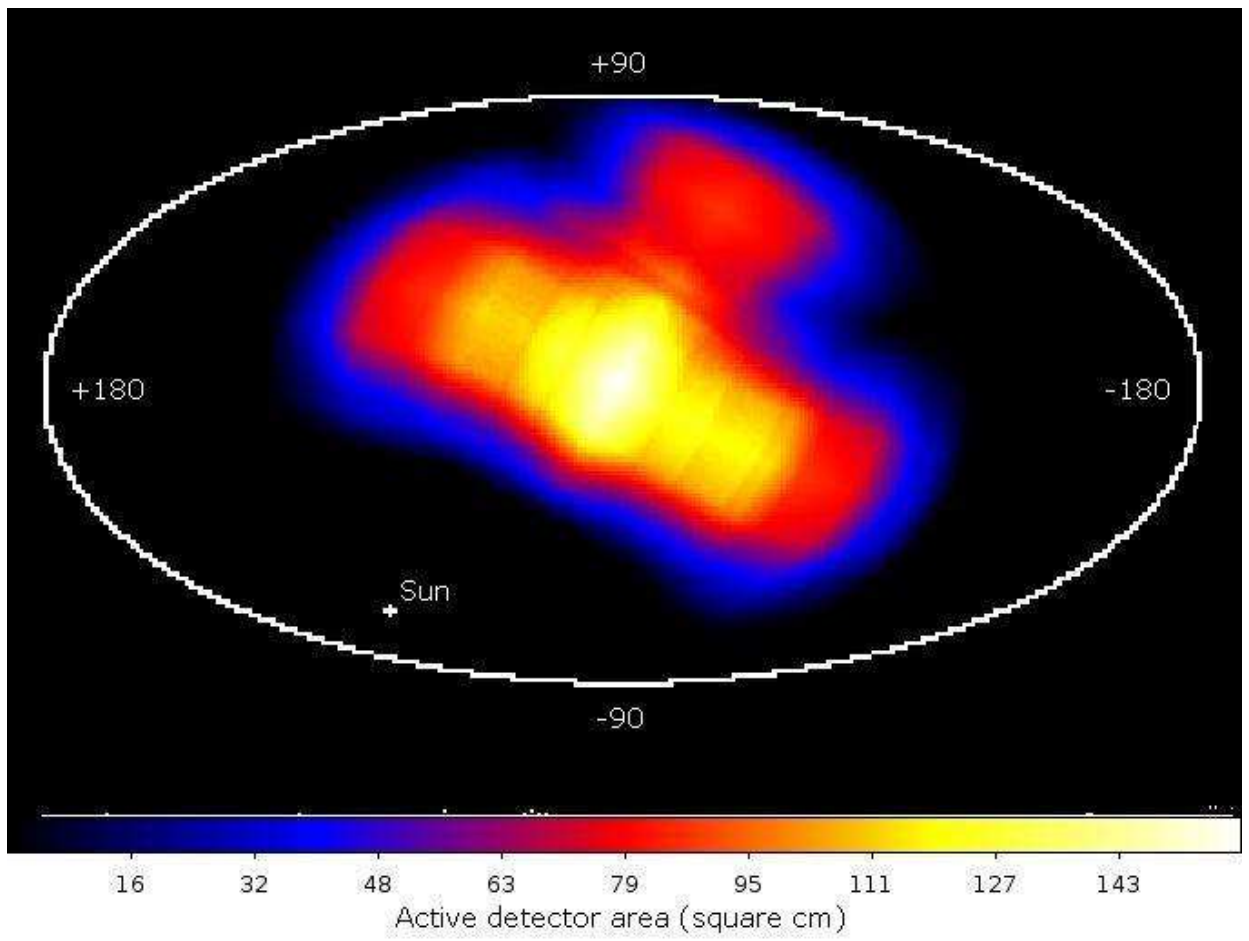

Figure 5. The projected effective area of the full WFM experiment, in Galactic coordinates.

Due to its large field of view, energy bandpass and imaging properties, the WFM is anticipated to detect and localize a large number of gamma-ray bursts and other fast transients every year. One of the 5 units is oriented in the anti-Sun direction, where the detected transients can be more easily followed-up by ground-based optical telescopes. To assist this type of observation, the onboard data processing envisages a triggering and imaging system to calculate the coordinates of the transient event in nearly real-time and distribute them world-wide through a VHF transmission system. In fact, due to telemetry limitations, the WFM will normally work by integrating detector images onboard every 5 minutes (integration time is programmable), in different energy bands. For short-duration events, the onboard triggering system will enable event-by-event data storage into the mass memory, for an approximate duration of 300 seconds. The maximum sustainable rate of triggers is one per orbit, thus allowing a large number of type I X-ray bursts to be detected as well. The WFM will be allowed to use the telemetry bandwidth left available from the LAD and will also transmit full event information for (part of) the WFM outside triggering events.

---

[1] The extended energy range 50-80 keV is for the purpose of LAD-background monitoring, not for the standard WFM science operation which is limited to 50 keV.

An extensive description of the WFM experiment may be found in Brandt et al. 2012. The main scientific requirements are summarized in Table 4.

Table 4. The main scientific requirements of the WFM.

| **Parameter** | **Value** |
|---|---|
| Energy Range | 2-50 keV primary<br>50-80 keV extended |
| Active Detector Area | 1820 $cm^2$ |
| Energy Resolution FWHM | 300 eV @ 6 keV |
| Field of View (Zero Response) | 180°x90° + 90°x90° |
| Angular Resolution | 5’ x 5’ |
| Point Source Location Accuracy (10-σ) | 1’ x 1’ |
| Sensitivity (5-σ, on-axis) | |
| Galactic Center, 3s | 270 mCrab |
| Galactic Center, 1 day | 2.1 mCrab |
| Standard Mode | 5-min energy resolved images |
| Trigger Mode | Event-by-event (10µs resolution)<br>Real-time downlink of transient coordinates |

## 4. MISSION FEATURES

The unprecedentedly large area of the LAD drives the design of the LOFT mission. As we have shown, the deployment in space of ~18 $m^2$ overall surface has been solved by using a set of 6 panels (the number of panels is not a requirement, and is open for optimization). In the preliminary design included in the original M3 proposal, satellite mass estimates were compatible with the small-class VEGA launcher. Later studies by the CDF Team at ESA ended up with a mass estimate (including all margins) of 2200 kg, already marginal (by ~10%) for the current estimate of the VEGA injection capabilities into a low-Earth equatorial orbit (in turn reduced with respect to the relevant User Manual, as a margin for the current stage of the VEGA development). In addition, the uncontrolled re-entry risk was estimated as being above threshold, requiring the addition of a re-entry control engine, and further increasing the mass budget to above 2400 kg. The overall conclusion of the preliminary ESA CDF study was that to maintain compatibility with VEGA the LAD area should be reduced by at least 25%. This was considered to be unacceptable by the LOFT Study Science Team, leading us to switch to the medium-class Soyuz launcher.

The use of a Soyuz for LOFT offers a large margin in volume and a huge margin in mass. For this reason, the cost increase for a more powerful launcher can be partially recovered by a significant de-risking on the satellite design, which in some cases could use off-the-shelf items. A further advantage of the margins provided by the Soyuz lies in the choice of the orbit. The major challenge for end-of-life spectral performance of the SDDs is radiation damage induced by the trapped protons in the South-Atlantic Anomaly. This is smaller at lower inclinations and altitudes. Low altitudes require periodic orbit up-lifting due to atmospheric dragging, then a dedicated engine and fuel. Low inclinations require correction at the launch phase (the launch base, Kourou, has a declination of ~5°), then additional fuel. Both of these orbital optimizations would have been unavailable with a VEGA launcher. The availability of a low-inclination (<2°), low-altitude (~550 km) orbit is also very favorable from the point of view of the particle-induced instrumental background, although this is not the dominant component in the LAD.

The attitude and orbit control system (AOCS) for LOFT is requested to be 3-axis stabilized. However, the huge counting statistics provided by LOFT on bright sources will require an accurate control on the possible systematic uncertainties introduced by the LAD response stability, as due to the AOCS or other sources of instability (e.g., thermal). This issue has been studied in detail by the LOFT Science Team in order to generate a sensible and focused requirement for the system. The requirement was studied as a function of the frequency, taking into account the expected astrophysical signal in each frequency range, as well as the discovery space open by the LAD counting statistics. The result is a requirement (rms) going from 2% stability (per decade) below 0.01 Hz, to 0.2% in 0.01-1 Hz, to 0.02% (per octave) in 1 Hz – 1 kHz. The approach to meeting such a requirement is to combine the angular response of the collimator (ideally triangular, but in practice smoothed by the element-to-element misalignment) with the AOCS parameters: the absolute pointing error should be such that (~arcminute) the source is always in the most flat, central part of the LAD field of view, whereas the relative pointing error should be small enough (a few to 15 arcseconds, depending on frequency) that the target source "samples" a region of the LAD response varying less than the required amplitude stability, for the given frequency range. It is important to note that such requirements do not directly apply to the AOCS parameter, since the attitude control frequencies will be significantly smoothed by the satellite structure, before being transferred to the LAD panels.

In Section 2 we have outlined the driving science objectives of the LOFT mission. The total observing time needed to satisfy all the top-level science goals amounts to 40% of the anticipated net observing time in 4 years. However the duration of the mission is driven not by the integrated observing time, but rather by the probability of detecting the most interesting events, such as the outburst of a black-hole transient. This means that a total of 30 Ms of observing time is available to exploit LOFT's capabilities as a general observatory. The Observatory Science Working Group of the LOFT Consortium is currently studying the best science cases, to make sure that the instrumental and mission parameters are compatible. As an example, this led to inclusion of the onboard (and on-ground) system for a rapid dissemination of the coordinates of gamma-ray bursts and transients.

In terms of enhancing LOFT's performance and flexibility, the team is working on two main items: increasing the LAD Field of Regard (FoR, the fraction of the sky accessible at any time, while satisfying all the requirements) and optimizing the background level and control. Extending the FoR beyond the required 50% implies the ability to control the LAD temperature over a wider range of Sun aspect angles. By design, the LAD uses a passive cooling system. The SDDs must operate at low temperature (below ~ -15 °C) to guarantee spectral performance at the end of life (after 4 years of radiation damage). The challenge when increasing the FoR is to maintain the operating temperature within the requirement, while exposing the detectors to widely different orientations to the Sun. Preliminary analysis by the ESA CDF Team shows that the extension of the FoR into the anti-Sun direction has the highest probability of being feasible (and this is another reason to orient the 5$^{th}$ WFM unit in the anti-Sun region). The work is still in progress. However, for some science objectives the full energy resolution is not a main requirement, as they do not deal with narrow spectral features. For these cases an additional requirement has been generated for the satellite to be able to point outside of the FoR while accepting a controlled degree of degradation of the spectral performance (typically a factor 1.5 worse than the requirement) due to increased temperature. This will allow us to extend significantly the flexibility of the mission, especially for ToOs. Anticipated ToOs related to the top-level science goals do not require fast repointing (1-2 hours or less) of the LAD but can be processed with reaction times of the order of several hours to 1-2 days. For this reason, the ToO reaction time requirement to the LOFT ground segment is 12 hours during working hours and 24 hours otherwise. These times actually correspond to requirements, although usually the type of ground segment foreseen for LOFT would react much more quickly.

As discussed, the use of a capillary plate collimator for the LAD is one of the two enabling technologies necessary to achieve 10 $m^2$ of effective area, but at the price of a background that is higher than previous missions (a few mCrab at 2 keV to 10 mCrab at 10 keV). Design optimizations are being studied to minimize the background level. However, for most if not all the science observations the real limit is not the absolute level of background, rather knowledge of its level and the systematic uncertainty after its subtraction. In this respect, the nature of the LAD background is far more favorable than previous missions. The LAD background is dominated (>90%) by the counts induced by the hard X-ray photons of the cosmic X-ray background and Earth albedo "leaking" through the collimator. Both sources are stable in time and they are expected not to induce any background variability. However, the position and orientation of the LAD experiment in this stable "radiation environment" (CXB for 70% of the solid angle and Earth albedo for 30%) varies along the orbit and with the attitude. As the two photon sources have different intensity and spectra, the relative movement of the LAD in their environment is detected as a variation. We simulated this effect and found that the maximum anticipated variation is ~20%. This should be compared to variability by a factor ~2-3 for experiments dominated by particle-induced background. The LAD background variability is therefore low-amplitude and low-frequency (orbit) and highly predictable being due to geometric effects only. To improve further on this, the LAD will include a sub-set of detectors equipped with a "blocked" version of the collimator, with exactly the same stopping power as the standard LAD collimator, but without apertures. These "background detectors" will detect counts from all background sources except for the aperture background and point sources in the FoV. The rates measured from the blocked detectors can then be used as continuous benchmarks to the background modeling. Devoting one Module to this task (or the equivalent 16 detectors in different Modules) the available counting statistics will enable fitting an orbital modulation with an accuracy of 0.3% per orbit and better on longer timescales. Together with the analytic modeling, this is expected to provide a systematic uncertainty on the background subtraction better than 0.5%.

The huge throughput of the LOFT mission requires a very efficient telemetry downlink system. The baseline is to use 2 ground stations (Kourou and Malindi) in the X band. The available data transfer per orbit is >7 Gbits. This matches the requirement of transmitting full event information for sources with intensity up to 0.5 Crab. For transient bright events (e.g., flaring) the excess telemetry will be stored onboard and gradually transmitted in the successive orbits. For sources with persistent intensity brighter than 0.5 Crab, the observing plan will optimize the observation of bright and dim sources (e.g., AGNs) to allow for the onboard storage and late transmission of the excess telemetry. In the most extreme cases, binned modes are planned, optimized to the specific science goals (i.e., privileging either time or spectral resolution), similar to the strategy successfully adopted by RXTE/PCA.

## 5. CONCLUSIONS

The Assessment Study of the LOFT ESA M3 mission candidate is being carried out by a consortium of European institutes for the payload instrumentation, and by the ESA Study Team and its industrial contractors for the spacecraft and system aspects. The ongoing study has so far confirmed the feasibility of the mission within the programmatic constraints of the ESA M3 call, with the same mission profile identified in the original proposal. The only significant changes are in an evolution of the Wide Field Monitor to a larger instrument with higher performance, and the switch from a VEGA baseline launcher to a Soyuz. On the payload side, all study and technology development activities are in place to meet the requirement of reaching a technology readiness level ≥5 by the end of 2014. A recent decision by ESA plans an extension of the assessment study until the completion of a full phase A by the end of 2013, when a down-selection to a single M3 mission will be carried out by the ESA Advisory Structure. The current ESA M3 baseline envisages a launch opportunity in 2022-24.

## ACKNOWLEDGEMENTS

The LOFT Consortium is grateful to the ESA Study Team and CDF Team for their professional and effective support to the study of the mission.

The Italian team is grateful for support by ASI (under contract I/021/12/0-186/12), INAF and INFN. The work of the MSSL and Leicester groups is supported by the UK Space Agency. The work of SRON is funded by the Dutch national science foundation (NWO). The work of the group at the University of Geneva is supported by the Swiss Space Office. The work of IAAT on LOFT is supported by the Germany's national research center for aeronautics and space DRL. The work of the IRAP group is supported by the French Space Agency. LOFT work at ECAP is supported by DLR under grant number 50 00 1111.